\documentstyle[twoside,fleqn,epsfig,amssymb,amsmath,espcrc2,graphicx]{article}

\newcommand{\msbar}{{\overline{\rm MS}}}
\newcommand{\bea}{\begin{eqnarray}}
\newcommand{\eea}{\end{eqnarray}}
\newcommand{\beq}{\begin{equation}}
\newcommand{\eeq}{\end{equation}}

\newcommand{\pdir}{p\kern -5.2pt\raise 0.2ex\hbox {/}}
\newcommand{\vdir}{v\kern -5.75pt\raise 0.15ex\hbox {/}}
\newcommand{\kdir}{k\kern -5.75pt\raise 0.15ex\hbox {/}}
\newcommand{\epsdir}{\epsilon\kern -5.0pt\raise 0.15ex\hbox {/}}
\newcommand{\bvdir}{\bar{v}\kern -5.75pt\raise 0.15ex\hbox {/}}
\newcommand{\Ddir}{D\kern -7.75pt\raise 0.20ex\hbox {/}}
\newcommand{\ldir}{l\kern -5.0pt\raise 0.2ex\hbox{/}}
\newcommand{\varepsdir}{\varepsilon\kern -5.5pt\raise 0.15ex\hbox{/}}

\newcommand{\bbar}{B^0-\bar B^0}
\newcommand{\ri}{{\rm RI/MOM}}

\title{Combined Relativistic and static analysis for all $\Delta B=2$ operators}

\author{D.~Be\'cirevi\'c\address{Dip. di Fisica, 
Universit\`a ``La Sapienza" and INFN-Roma, P.le A. Moro 2, I-00185 Rome,
Italy}, 
V.~Gim\'enez\address{Dep.~de F\'{\i}s.Te\'orica and IFIC, Univ.~de Valencia, Dr.~Moliner 50, E-46100, Burjassot, Valencia,
Spain}, 
G.~Martinelli$^{\rm a}$, M.~Papinutto\address{Dip. di Fisica, Univ. di Pisa 
and INFN - Pisa,
Via Buonarroti 2, I-56100 Pisa, Italy},
J.~Reyes$^{\rm a,\rm b}$\thanks{
Talk given by Juan Reyes.}}
\begin{document}
\newcommand{\sze}{\small}

\begin{abstract}
We analyse matrix elements of $\Delta B=2$ operators by combining QCD
results with the ones obtained in the  static limit of HQET.
The matching of all the QCD operators to HQET is made 
at NLO order. To do that we
have to include the anomalous dimension matrix up to two loops, both in QCD and
HQET, and the one loop matching for all the $\Delta B=2$ operators. The matrix elements of these operators
are relevant for the prediction of the $B-\bar B$ mixing, $B_s$ meson width
difference and supersymmetric effects in $\Delta B=2$ transitions. 
\end{abstract}
\maketitle

\section{INTRODUCTION}
Matrix elements of $\Delta B=2$ operators crucially enter 
theoretical determinations of important phenomenological quantities.
Due to the high mass of the $b$ quark and to the present computing  power, it is
impossible to simulate $b$ quarks directly on the lattice. Two
approaches 
have been employed to overcome this problem:
\begin{itemize}
\item Perform relativistic simulations with several masses around the $c$-quark
mass and extrapolate the results to physical $b$-quark mass.
\item By means of an effective theory, the heavy degrees of
freedoms can be integrated out and only the light modes are
simulated. So far, HQET and NRQCD have been employed. In HQET an
expansion in $\Lambda^{QCD}/m_b$ is made whereas in  NRQCD the limit $v\ll 1$ is 
taking, being $v$ the heavy quark velocity.
\end{itemize}

In this work we have combined, for the first time, full QCD and HQET
simulations to compute the matrix elements of $4$-quark
$\Delta B=2$ operators. To do so appropriately, the computation of the matching
coefficients of the operators between QCD and HQET at next-to-leading order
(NLO) accuracy is needed. 

We have considered the following basis of $4$-quark $\Delta B=2$ operators:
\bea
 \label{basisc}
O_1 &=& \ \bar b^i \gamma_\mu (1- \gamma_{5} )  q^i \,
 \bar b^j  \gamma_\mu (1- \gamma_{5} ) q^j \,  , 
  \nonumber \\
O_2 &=& \ \bar b^i  (1- \gamma_{5} ) q^i \,
\bar b^j  (1 - \gamma_{5} )  q^j \, ,  \nonumber  \\
O_3&=& \ \bar b^i  (1- \gamma_{5} ) q^j \,
 \bar b^j (1 -  \gamma_{5} ) q^i \, ,  \\
O_4 &=& \ \bar b^i  (1- \gamma_{5} )  q^i \,
 \bar b^j   (1+ \gamma_{5} ) q^j \,  ,  \nonumber \\
O_5 &=& \ \bar b^i  (1- \gamma_{5} )  q^j \,
 \bar b^j   (1+ \gamma_{5} ) q^i \,  ,  \nonumber 
 \eea        
with $i$, $j$ colour indices, and $q$ stands for 
either $d$- or $s$- light quark flavour. 
The first of the above operators enters the Standard Model (SM) description of the $\bbar$ mixing
amplitude \cite{buras1}, whereas 
$O_2$ and $O_3$ are relevant for the  
SM estimates of the relative width difference in the neutral $B$-meson system, 
$\left( \Delta \Gamma/\Gamma\right)_{B_s}$~\cite{beneke}. The matrix
elements of all 
operators parametrize supersymmetric effects in $\Delta B=2$ transitions
\cite{luca}.

It is usual to express the matrix elements of the operators~(\ref{basisc}) 
in terms of the so-called B-parameters, which are introduced as a 
measure of the deviation from the vacuum saturation approximation (VSA),
namely~\cite{luca,allton}, 
\bea
 \label{params}
\langle \bar B^0_q \vert \hat O_1(\mu) \vert   B^0_q \rangle  &=& b_1 \, m_{B_q}^2  f_{B_q}^2
B_1(\mu) \ ,  \nonumber\\
\langle \bar B^0_q \vert \hat O_i(\mu) \vert   B^0_q \rangle  &=& b_i \,
\chi\,
m_{B_q}^2  f_{B_q}^2 B_i(\mu),\\
&& 2\le i\le5\nonumber 
\eea                                                                            
with $\vec b=\{8/3,\;-5/3,\;1/3,\;2,\;2/3\}$ and
$\chi \equiv m_{B_q}^2/ (m_b(\mu) + m_q(\mu) )^2$.
The hat symbol denotes operators renormalized in some 
renormalization scheme at the scale $\mu$. 
To determine the values of the B-parameters $B_{1-5}(\mu)$, we
have combined the results of a (quenched) QCD numerical simulation 
on the lattice with 3 values of the heavy quark mass, in the range of heavy-light
pseudoscalar  masses $m_P\in (1.7, 2.4)$~GeV,  with HQET results in order to   
constrain the extrapolation towards the physical point, 
$m_{B_{s/d}}$. This extrapolation is guided by the Heavy Quark Scaling Laws
(HQSL)
which are the propertly ones of HQET. Therefore, to use the HQSL we have to
match the $B$-parameters obtained in QCD onto the HQET ones. 
We refer the reader to ref. \cite{paper} where this issue is explained in
great
detail.
\section{SIMULATION DETAILS}
We enumerate, briefly, the main elements of our lattice 
simulations:
\begin{itemize}
\item
For full QCD: the details can be found in refs.~\cite{ape1,ape2,ape3}.
The simulation is performed in a lattice of the size $24^3\times 48$, at $\beta
= 6.2$ corresponding to $a^{-1}=2.7$ GeV, with the non-perturbatively improved Wilson action. 
The number of gauge configurations is 200, with 3 values of the heavy and 3 values of
the light quark masses, corresponding to the Wilson hopping 
parameters: $\kappa_q \in \{ 0.1344, 0,1349, 0.1352\}$, and 
$\kappa_Q \in \{ 0.125, 0,122, 0.119\}$. 
\item For HQET: the details are in ref.~\cite{gimmar}. The simulation is
performed in a lattice of the size $24^3\times 40$, at $\beta 
= 6.0$  corresponding to $a^{-1}=2.0$ GeV, with the tree level Clover improved Wilson action. 
The number of gauge configurations is 600, with 3 values of
the light quark masses, corresponding to the Wilson hopping 
parameters $\kappa_q \in\{ 0.1425, 0.1432, 0.1440\}$.
\end{itemize}
\section{EXTRAPOLATION TO THE B-MESON}
Once the B-parameters of the renormalized operators, in QCD and HQET, are
obtained from lattice simulations in some scheme and at some scale, the matching
between the two theories reads: 
\bea \label{match1}
W^T_{QCD}[m_P,\mu]^{-1}\cdot \vec B (m_P,\mu) =\phantom{C}\hspace{.2cm}&&\nonumber\\
C(m_P)\cdot
 W^T_{HQET}[m_P,\mu]^{-1}
\cdot\vec{\widetilde B} (\mu) 
\nonumber\\ 
\ +\ {\cal
O}\left({1\over m_{P}}\right)+\dots \phantom{C(m_P)\cdot
 W^T_{HQ}}\hspace{.2cm}&& 
\eea
where $ W^T_{QCD}[\mu_2, \mu_1]^{-1}$ is the matrix encoding 
the full QCD evolution from a scale $\mu_1$ to $\mu_2$ of all five $\Delta B=2$ 
B-parameters ,
likewise for $ W^T_{HQET}[\mu_2, \mu_1]^{-1}$ in HQET. The matrix $C$ is the
matching 
matrix of the B-parameters between QCD and HQET. These matrices 
are specified in \cite{paper} . $\vec B (m_P,\mu)$ is a five-component vector
which collects all  five $\Delta B=2$ 
B-parameters renormalized at the scale $\mu$ simulated with a heavy quark mass
corresponding to a pseudescalar meson mass of $m_P$. $\vec{\widetilde
B}(\mu)$  is the corresponding vector in HQET. 

In order to account for the logarithm dependence in eq.~(\ref{match1})
we put the HQET evolution matrix in the l.h.s. and construct the quantity
$\vec \Phi(m_P,\mu)$:
\bea \label{match2}
\vec \Phi (m_P,\mu) =\phantom{ W^T_{HQET}[m_P,\mu] \cdot C^{-1}(m_P)\cdot
W^T} \hspace{-.1cm}&&\nonumber\\
 W^T_{HQET}[m_P,\mu] \cdot C^{-1}(m_P)\cdot
W^T_{QCD}[m_P,\mu]^{-1}\hspace{-.1cm}&&\nonumber\\
\cdot \vec B (m_P,\mu)\phantom{ W^T_{HQET}[m_P,\mu] \cdot C^{-1}(m_P)\cdot
W^T_{QC}} &&
\eea
which can be fit either freely as
\bea \label{fit10}
\vec \Phi (m_P,\mu) = \vec a_0(\mu) +  {\vec a_1(\mu) \over m_P}\,,
\eea
where  $\vec a_0(\mu)$ and $\vec a_1(\mu)$ are the fit parameters,  
or by constraining it by the static HQET results, {\it i.e.}
\bea \label{fit20}
\vec \Phi (m_P,\mu) = \vec a_0^\prime(\mu) + {\vec a_1^\prime(\mu) \over m_P}+ 
{\vec a_2^\prime(\mu) \over m_P^2}\;,
\eea
where the coefficient $\vec a_0^\prime(\mu)$ is completely constrained by the
static value, 
 $\vec {\widetilde B}(\mu)$, so that one can probe the term 
${\cal O}(1/m_P^2)$. In figure 1 we show the two extrapolation of eqs.
\ref{fit10} and \ref{fit20} 
for the first three B-parameters. In \cite{paper} the analogous plots for
$\Phi_4$ and $\Phi_5$ can be found.
As a result, we obtain the HQET values of
the $B$-parameters, {\it i.e.} $\vec \Phi (m_{B_{s/d}},\mu)$, which are 
then to be matched back onto their QCD counterparts. The final results in the
constrained case are presented in table 1, where the first error is the
statistical one and the second the systematic error, due to the
uncertainty in the lattice renormalization constants, both in QCD and HQET
(see ref. \cite{paper} for more details).
\begin{figure}[t!]
\begin{center}
\vspace{0cm}
\includegraphics[bb=4.9cm 2cm 13cm 20cm,
width=4.1cm]{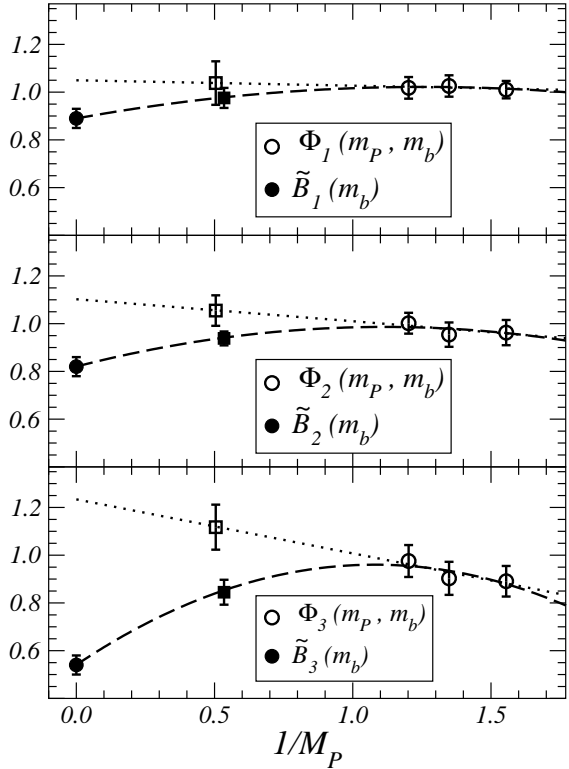}  \\

\caption{\label{fig33}{\sl \small Extrapolation to the physical 
$B_d$ meson mass (squared symbols) in the inverse heavy meson mass. 
The dotted line corresponds to the unconstrained linear extrapolation for each
of the components 
$\Phi_i(m_{P},m_b)$ from our data (empty circles) to $
\Phi_i(m_{B_d},m_b)$ (empty square). The result
of the constrained 
extrapolation (filled squares) by the static HQET B-parameters (filled circles)
is marked by the dashed line.}}
\end{center}
\end{figure}
\begin{table}[t!] 
\caption{\label{tab0}
\small{\sl $B$-parameters defined in
eq.~(\ref{params}), renormalized at $\mu = m_b = 4.6$~GeV and in two 
renormalization schemes: $\ri$ and $\msbar$ of ref.~\cite{buras2}.}}
\begin{center} 
\begin{tabular}{ccc} 
\hline
\mbox{}&\multicolumn{2}{c}{\sl Scheme}\\\hline
\mbox{}&  RI/MOM  &  $\msbar$(NDR)~\cite{buras2}   
  \\   \hline \hline 
 $B_1^{(d)}(m_b)$ & $0.87(4)\left({}^{+5}_{-4}\right)$& $0.87(4)\left({}^{+5}_{-4}\right)$  \\ 
 $B_2^{(d)}(m_b)$ & $0.82(3)(4)$ & $0.79(2)(4)$ \\ 
 $B_3^{(d)}(m_b)$ & $1.02(6)(9)$ & $0.92(6)(8)$  \\ 
 $B_4^{(d)}(m_b)$ & $1.16(3)\left({}^{+5}_{-7}\right)$ 
&  $1.15(3)\left({}^{+5}_{-7}\right)$\\ 
 $B_5^{(d)}(m_b)$ & $1.91(4)\left({}^{+22}_{-7}\right)$ & $1.72(4)\left({}^{+20}_{-6}\right)$
  \\ \hline
 $B_1^{(s)}(m_b)$ & $0.86(2)\left({}^{+5}_{-4}\right)$ & 
$0.87(2)\left({}^{+5}_{-4}\right)$
\\ 
 $B_2^{(s)}(m_b)$ & $0.83(2)(4)$ & $0.80(1)(4)$ \\ 
 $B_3^{(s)}(m_b)$ & $1.03(4)(9)$ & $0.93(3)(8)$ \\ 
 $B_4^{(s)}(m_b)$ & $1.17(2)\left({}^{+5}_{-7}\right)$ &  $1.16(2)\left({}^{+5}_{-7}\right)$ \\ 
 $B_5^{(s)}(m_b)$ & $1.94(3)\left({}^{+23}_{-7}\right)$ & $1.75(3)\left({}^{+21}_{-6}\right)$  \\ \hline
\end{tabular} 
\end{center}
\end{table}

This work has been supported by the 
European Community's Human potential programme under HPRN-CT-2000-00145
Hadrons/LatticeQCD.

\end{document}